\documentclass[12pt,a4paper]{article}

\usepackage{amssymb}
\usepackage{amsmath, epsfig, color, subfigure,bm}
\usepackage{fullpage}
\usepackage{color}

\newcommand{\be}{\begin{equation}}
\newcommand{\en}{\end{equation}}

\renewcommand{\vec}[1]{\boldsymbol{#1}}

\begin{document}

\title{\bf{At least three invariants are necessary to model the mechanical response of incompressible, transversely isotropic materials} }

\author{
M. Destrade$^{ab}$,  B. Mac Donald$^c$, J.G. Murphy$^{ca}$, G. Saccomandi$^d$ \\[10pt]
\emph{$^a$School of Mathematics,  Statistics, and Applied Mathematics,} \\
\emph{National University of Ireland Galway, University Road, Galway, Ireland.} \\[10pt]
\emph{$^b$School of Mechanical and Materials Engineering,} \\
\emph{University College Dublin, Belfield, Dublin 4, Ireland.} \\[10pt]
\emph{$^c$Centre for Medical Engineering Research,} \\
\emph{Dublin City University, Glasnevin, Dublin 9, Ireland.} \\[10pt]
\emph{$^d$Dipartimento di Ingegneria Industriale,}  \\
\emph{Universit\`{a} degli Studi di Perugia,  06125 Perugia, Italy.}
 }

\date{}
\maketitle

\begin{abstract}
The modelling of off-axis simple tension experiments on transversely isotropic nonlinearly elastic materials is considered. A testing protocol is proposed where normal force is applied to one edge of a rectangular specimen with the opposite edge allowed to move laterally but constrained so that no vertical displacement is allowed. Numerical simulations suggest that this deformation is likely to remain substantially homogeneous throughout the specimen for moderate deformations. It is therefore further proposed that such tests can be modelled adequately as a homogenous deformation consisting of a triaxial stretch accompanied by a simple shear. Thus the proposed test should be a viable alternative to the standard biaxial tests currently used as material characterisation tests for transversely isotropic materials in general and, in particular, for soft, biological tissue. A consequence of the analysis is a kinematical universal relation for off-axis testing that results when the strain-energy function is assumed to be a function of only one isotropic and one anisotropic invariant, as is typically the case. The universal relation provides a simple test of this assumption, which is usually made for mathematical convenience. Numerical simulations also suggest that this universal relation is unlikely to agree with experimental data and therefore that at least three invariants are necessary to fully capture the mechanical response of transversely isotropic materials.
\end{abstract}


\section{Introduction \label{Introduction}}


The recent resurgence in interest in the modelling of the mechanical response of incompressible, transversely isotropic, nonlinearly elastic materials is primarily because there are many examples of biological, soft tissue reinforced with bundles of fibres that have an approximate single preferred direction, most notably skeletal muscles, ligaments and tendons. Developed mainly by Rivlin and co-workers  \cite{Rivpap},  the phenomenological constitutive theory for such materials was originally used to model elastomers reinforced with steel cords and it is a happy coincidence that such an elegant, rational theory can be applied to some of the fundamental modelling problems in biomechanics. 

If soft tissue is assumed non-dissipative, as is commonly the case, then its mechanical response is determined completely by the corresponding potential function, called the strain-energy function in mechanics. For incompressible, transversely isotropic materials, the general strain-energy function is an arbitrary function of four scalar invariants (see, for example, Spencer \cite{Spen}). These invariants are functionals of a strain tensor, with the left Cauchy-Green strain being the measure of choice in biomechanics, and the direction of the fibres in the undeformed configuration. Unfortunately, the corresponding stress-strain relation is algebraically complex (see \eqref{ssa}) and, to facilitate analysis, simplifying assumptions are, and usually must, be made. By far the most common assumption is to assume that the strain-energy is a function only of the two invariants $I_1,I_4$, where $I_1$ is the trace of the Cauchy-Green strain tensor and $I_4$ is the square of the fibre stretch. As already mentioned, this choice is usually made on the basis of mathematical convenience alone, as a cursory examination of \eqref{ssa} reveals the more complicated terms in the stress-strain relation are eliminated as a result. This choice of the $\left(I_1,I_4\right)$ pair as the basis for the strain-energy function is the central theme here. It will be critically evaluated by examining the mechanical  response of transversely isotropic materials in the simplest material characterisation test: the simple tension test.
  
In contrast to the situation for isotropic materials, the development of protocols and methods for the simple tensile testing of \emph{anisotropic} materials is still on-going despite its long history (see, for example, Pagano and Halpin \cite{PaH}). Simple tension testing when the direction of anisotropy is oblique to the direction of the applied force is referred to as `off-axis testing'.  It has long being recognised that the standard rigid grips of most tensile testing machines induce shear forces and bending moments in the test specimens during off-axis testing, resulting in large stress concentrations and inhomogeneity in the test samples. This coupling of simple tension with shearing forces in off-axis testing of fibre-reinforced materials has, in fact, been exploited to characterise intralaminar shear (see Mar$\acute{\i}$n \emph{et al.} \cite{Marin}). References to the different methods employed to reduce stress concentrations and inhomogeneity of the tested sample in off-axis testing can be found in Mar$\acute{\i}$n \emph{et al.} \cite{Marin} and Xiao \emph{et al.} \cite{Xea}. 

Heretofore the vast majority of off-axis testing has involved epoxy composites and is therefore concerned with the determination of material constants within the context of the linear theory. Here the focus is on off-axis testing within the nonlinear regime and a new method of testing unidirectional composites is considered. Specifically it is proposed here that the coupling of shear and simple tension in off-axis testing be fully recognised and that the specimen, classically fully constrained along one of its edges, be allowed to move laterally there. A combination of tri-axial stretch and simple shear is therefore proposed to model the resulting deformation. This stretch/shear combination has previously been studied for \emph{isotropic} materials by, amongst others, Moon and Truesdell \cite{Moon}, Rajagopal and Wineman \cite{RaW} and, more recently, by Mihai and Goriely \cite{Goriely} and by Destrade \emph{et al.} \cite{Dea}.

The main objective here is to consider the validity of assuming pairs of invariants as the sole arguments of the strain-energy function. To this end, numerical simulations based on the Finite Element Method (FEM) were performed. The commercial FEM programme ANSYS was used throughout. There is no direct implementation of a transversely isotropic material available in ANSYS. Instead fibre-reinforced nonlinearly elastic matrices were modelled using a structural modelling approach, although this structural model includes a phenomenological component, the neo-Hookean matrix. This matrix is  reinforced by much stiffer linearly elastic cords. On clamping one of the edges to restrict movement in the vertical direction, but allowing displacement in the lateral direction, a normal force is applied to the opposite edge. 

Two inferences can be drawn from an analysis of the output of the various simulations performed by varying the applied force and the fibre orientation. The first is that a homogenous deformation field consisting of a triaxial deformation accompanied by a simple shear seems a viable model of material behaviour for off-axis testing. This gives some support to the viability of the experimental method proposed here. The second is that the assumption that pairs of invariants are sufficient to capture the main features of the mechanical response of nonlinearly elastic, transversely isotropic materials is not compatible with our numerical simulations and therefore that \emph{at least three invariants are necessary to fully capture the mechanical response of transversely isotropic materials}. 
This is because if pairs of invariants are assumed, then a universal kinematical relation between the kinematical variables results from satisfaction of the boundary conditions associated with simple tension. An illustrative comparison is then made between the kinematical relation that results from the $\left(I_1,I_4\right)$ choice and the FEM results for an illustrative fibre orientation over a range of physiological strain. It will be shown that there is a fundamental incompatibility between the two sets of results.
 
Interpretation of this incompatibility will depend on which model is more likely to encapsulate the mechanical response of biological, soft tissue in the laboratory. Ultimately, of course, this question can only be resolved by conducting simple tension tests of the type proposed here but, faced with the lack of experimental data, it is our contention that the physically well-motivated structural model that is the basis of our FEM results is a better choice than a phenomenological model based primarily on mathematical convenience. Using Finite Element Analysis in this way to inform the constitutive modelling process seems a novel application of computational mechanics. 

The paper is organized as follows: after a section outlining the constitutive theory for transversely isotropic materials, the modelling of off-axis simple tension tests is discussed in Section \ref{st}, with a particular emphasis on the modelling of a class of materials often used in biomechanics. The results of our numerical experiments are then reported in Section \ref{num} and the consequences of these results for the modelling of transversely isotropic, soft tissue are discussed.

Although the assumed homogeneous deformation consisting of a simple shear superimposed upon a triaxial stretch seems a natural fit with off-axis testing of transversely isotropic materials, and seems supported by our numerical experiments, the semi-inverse approach adopted here is not without its limitations, similar to the problems associated with modelling simple shear for \emph{isotropic} materials (Rivlin \cite{Rivlin}, Gent \emph{et al.} \cite{Gent}, Horgan and Murphy \cite{HaM}). Specifically, stresses need to be applied to the inclined faces of the deformed test pieces in order to maintain the assumed combination of tri-axial stretch and simple shear. This is discussed in the final section.


\section{Transversely isotropic materials\label{Basic}}


Incompressible fibre-reinforced materials are considered from now on. 
We call $\vec x = x_i \vec e_i$ the coordinates in the current configuration $\mathcal B$ of a particle which was at $\vec X = X_\alpha \vec E_\alpha$ in the reference configuration $\mathcal B_r$. 
Here the orthonormal vectors $\vec E_\alpha$ are aligned with the edges of the test sample, which is assumed to be a cuboid of dimensions
\be
-A/2 \le X_1 \le A/2, \qquad 
-L/2 \le X_2 \le L/2, \qquad
-H/2\le X_3 \le H/2.
\en
We take the the orthonormal vectors ($\vec e_1, \vec e_2, \vec e_3$) to be aligned with ($\vec E_1, \vec E_2, \vec E_3$).
The sample is clamped at $X_2=\pm L/2$ and the clamps are attached to the cross-heads of a tensile machine. 
One of the clamps is allowed to slide freely in the $X_2$ direction. 

We assume that there exists a single preferred direction (along the unit vector $\vec A$, say) to which all reinforcing fibres are parallel (transverse isotropy) and that  the fibres are confined to the ($X_1,X_2$) plane in the undeformed configuration. Thus  
\be \label{aa}
\vec{A}=C \vec{E}_1+S \vec{E}_2,
 \qquad \text {where } C \equiv \cos\Phi, \quad S \equiv \sin\Phi,
 \en
 where $\Phi$ $(0 < \Phi < \pi/2)$ is the angle in the undeformed configuration between the fibers and the direction normal to the tensile force.

Let $\vec F \equiv \partial \vec x / \partial \vec X$ denote the deformation gradient tensor and $\vec B = \vec{FF}^T$, $\vec C = \vec F^T \vec F$ the left and right Cauchy-Green strain tensors, respectively.
For incompressible materials, $\det \vec F = 1$.
The general strain-energy function for incompressible, fibre-reinforced, hyperelastic materials has the form $W=W(I_1, I_2, I_4, I_5)$  (see Spencer \cite{Spen}), where $I_1$, $I_2$ are the first and second isotropic principal invariants:
\be
I_1 = \text{tr}\, \vec C, \qquad I_2 = \text{tr}\, \vec C^{-1},
\en
and $I_4$, $I_5$ are the anisotropic invariants,
\be
I_4= \vec{A \cdot CA}, \qquad I_5=\vec{A \cdot C}^2\vec{A}.
\en
 The corresponding Cauchy stress tensor $\vec{T}$ is given by  \cite{Spen}
\be \label{ssa}
\vec{T}=-p\vec{I}+2W_1\vec{B}-2W_2\vec{B}^{-1}+2W_4\vec{a}\otimes\vec{a}+2W_5(\vec{a}\otimes\vec{Ba}+\vec{Ba}\otimes\vec{a}),
\en
where $W_k \equiv \partial W/\partial I_k$, $p$ is a Lagrange multiplier introduced by the incompressibility constraint, $\vec I$ is the identity tensor and $\vec{a} \equiv \vec{FA}$.

Experience has shown that the technical challenges of analysing general transversely isotropic materials are formidable and indeed further evidence of this will be provided in later sections. To make progress simplifying assumptions need to be made. For transversely isotropic materials, it is usual to ignore the $I_2,I_5$  invariants and to adopt the  assumption that
\be \label{sepw}
W= W\left(I_1, I_4\right) \text{ only}. 
\en
Many strain-energy density functions used in biomechanics applications have this form (see, for example, Humphrey and Yin \cite{Hump1}, Humphrey \emph{et al.} \cite{Hump2}, Horgan and Murphy \cite{tor}, Wenk \emph{et al.} \cite{other}) and a much-used example is the so-called standard reinforcing material
\be \label{srm}
W(I_1, I_4)=\frac{\mu}{2} \left[I_1-3 +\gamma  (I_4-1)^2\right],
\en
where $\mu \, (>0)$ is the shear modulus of the neo-Hookean potential and $\gamma \, (>0)$ is a non-dimensional material constant that provides a measure of the strength of reinforcement in the fibre direction, with large values of this parameter typical for soft, biological tissue (see, for example, Ning \emph{et al.} \cite{Ning}, Destrade \emph{et al.} \cite{Dest}) . 
Another popular choice is the Gasser-Ogden-Holzapfel model \cite{holz1}
\be \label{holz}
W(I_1, I_4)=\frac{\mu}{2} \left(I_1-3\right) + \dfrac{k_1}{2k_2}\left[\text e^{  k_2(I_4-1)^2} -1\right],
\en
where $\mu$, $k_1$ and $k_2$ are positive constants, to be determined from experimental data.
Similarly, the extension of this strain-energy density to include dispersive effects for the fibers in \cite{holz2},  the so-called `HGO' model implemented in the finite element software ABAQUS, also belongs to the family \eqref{sepw}.


\section{Simple tension test: analytical solution} \label{st}


We focus on the general homogeneous field response generated by a tensile test where the tensile force occurs at an angle to the fibres. Hence we take the components $F_{i\alpha} = \partial x_i/\partial X_\alpha$ of the deformation gradient tensor to be constants.   
One clamp is allowed to slide in the direction of $\vec E_1= \vec e_1$ and the line elements that were parallel to the clamps in $\mathcal B_r$ remain parallel to the clamps in $\mathcal B$. 
In other words, the deformation takes the form
\be \label{def}
x_1 = F_{11}X_1 + F_{12}X_2, \qquad x_2=F_{22} X_2, \qquad x_3=F_{33}X_3.
\en
An illustrative example of this type of deformation is given in Figure \ref{figEXP} below.  Deformations of this form are a special case of the homogeneous deformations, with deformation gradient tensor
\be
[\vec{F}]_{i\alpha} = 
\begin{bmatrix}
F_{11} & F_{12} & 0
\\
F_{21} & F_{22} & 0
 \\
0 & 0 & F_{33}
 \end{bmatrix}
\en
considered by Holzapfel and Ogden \cite{HaO} who wished to clarify the extent to which biaxial testing can be used for determining the elastic properties of transversely isotropic materials. 

The deformation \eqref{def} can be decomposed as a tri-axial stretch accompanied by a simple shear, as can be seen from the following identifications in the ($\vec e_i \otimes \vec E_\alpha$) coordinate system,
\be
[\vec{F}]_{i\alpha} = 
\begin{bmatrix}
F_{11} & F_{12} & 0
\\
0 & F_{22} & 0
 \\
0 & 0 & F_{33}
 \end{bmatrix}
 = 
\begin{bmatrix}
\lambda_1 & \lambda_2 \kappa & 0
\\
0 & \lambda_2 & 0
 \\
0 & 0 & \lambda_3
 \end{bmatrix}
 = 
\begin{bmatrix}
1 & \kappa & 0
\\
0 & 1& 0
 \\
0 & 0 & 1
  \end{bmatrix}
\begin{bmatrix}
\lambda_1 & 0& 0
\\
0 & \lambda_2 & 0
 \\
0 & 0 & \lambda_3
 \end{bmatrix},
 \en
(see  \cite{Moon, RaW, Dea}  for isotropic materials).
Here $\lambda_1, \lambda_2, \lambda_3$ are positive constants, with $\lambda_3=(\lambda_1\lambda_2)^{-1}$ as a result of imposing the incompressibility constraint and $\kappa$ is the amount of shear in the $\vec E_2 = \vec e_2$ direction.
The data to be collected during those tensile tests are: $\lambda_1$, $\lambda_2$, $\kappa$ and $T_{22}$, the tensile Cauchy stress component. 
The stretches can be measured with two orthogonal LASER tracking devices and $\kappa$ by measuring the transverse displacement of the sliding clamp, see Fig.\ref{figEXP}. 
Alternatively, a Digital Correlation Imaging device can be used.
To measure $T_{22}$, the force is measured by a loadcell attached to a clamp and divided by $\lambda_1A\times \lambda_3 H = AH/\lambda_2$, the current cross-sectional area.
\begin{figure}
\begin{center} \epsfig{width=0.6\textwidth, figure=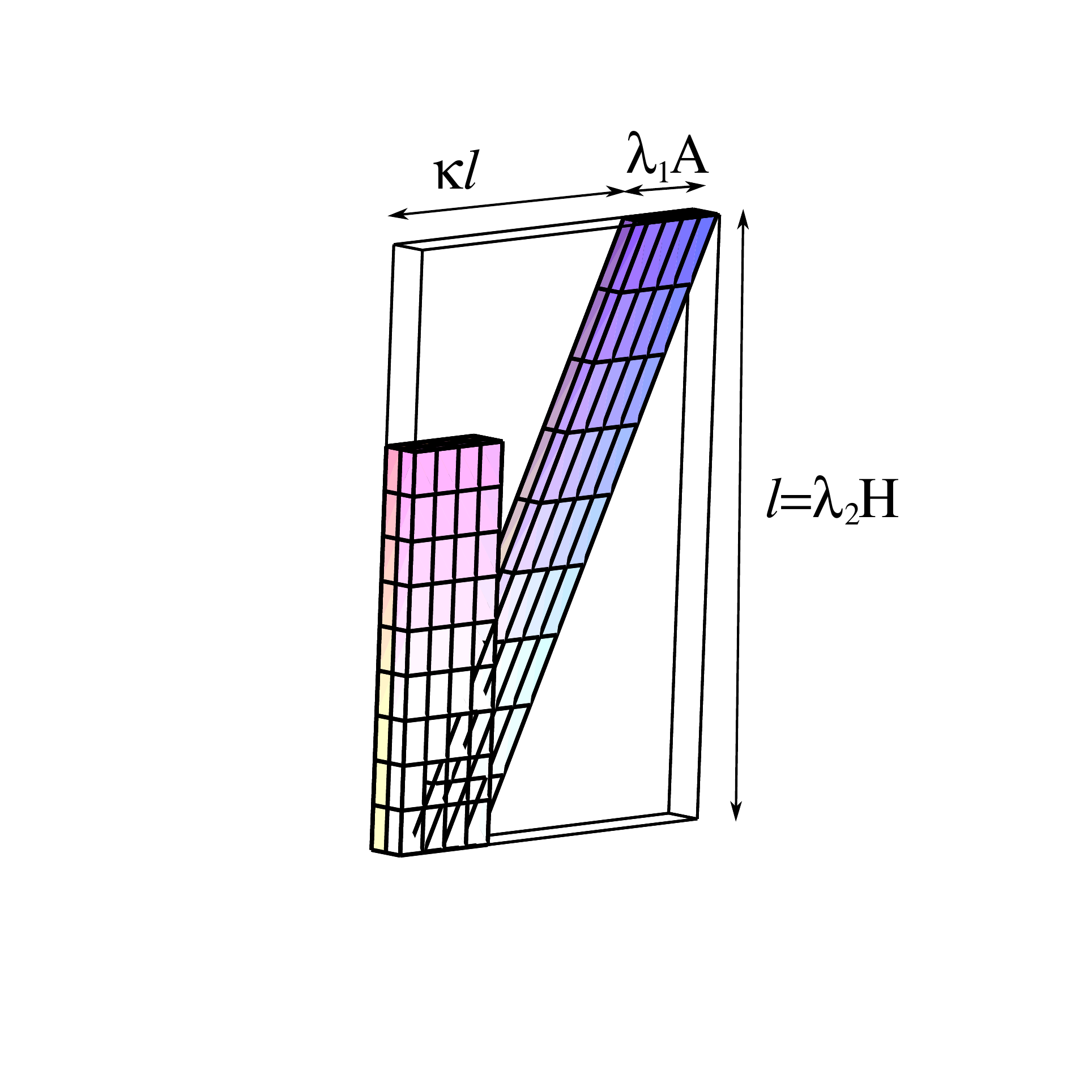}\end{center}
\caption{{\small A cuboid of hyperelastic material reinforced with one family of parallel fibers can deform homogeneously when subjected to a tensile force only, as a combination of triaxial stretch (with  stretch ratios $\lambda_i$) and simple shear (with amount of shear $\kappa$).}}
\label{figEXP}
\end{figure}

We now compute the components of the left and right Cauchy-Green deformation tensors as
 \be \label{matrices}
 [\vec{B}]_{ij} = 
\begin{bmatrix}
\lambda_1^2+\lambda_2^2 \kappa^2 &   \lambda_2^2 \kappa & 0
\\
\lambda_2^2 \kappa & \lambda_2^2 & 0
 \\
0 & 0 & \lambda_3^2
\end{bmatrix},
\qquad
 [\vec{C}]_{\alpha \beta} = 
\begin{bmatrix}
\lambda_1^2 &   \lambda_1 \lambda_2 \kappa & 0
\\
\lambda_1 \lambda_2 \kappa & \lambda_2^2(1+\kappa^2) & 0
 \\
0 & 0 & \lambda_3^2
\end{bmatrix},
\en
 in the ($\vec e_i \otimes \vec e_j$) and the ($\vec E_\alpha \otimes \vec E_\beta$) coordinate systems, respectively. 
The isotropic strain-invariants are given by \cite{RaW}
\be \label{invv}
I_1 = \lambda_1^{2} + \lambda_2^{2}\left(1+\kappa^2\right)+ \lambda_1^{-2}\lambda_2^{-2}, \qquad 
I_2= \lambda_1^{-2}\left(1+\kappa^2\right) + \lambda_2^{-2} + \lambda_1^{2}\lambda_2^{2},
\en
and the anisotropic invariants by
\begin{align} \label{i4}
& I_4 = \left(\lambda_1 C+\lambda_2\kappa S\right)^2+\lambda_2^2S^2, \notag \\
& I_5 =\lambda_1^2\left(\lambda_1^2 + \lambda_2^2 \kappa^2 \right)C^2 \notag \\
& \qquad \qquad + \lambda_2^2\left[\lambda_2^2+ \left(\lambda_1^2 + 2\lambda_2^2 \right)\kappa^2 + \lambda_2^2 \kappa^4 \right] S^2 +2 \lambda_1\lambda_2\left[\lambda_1^2+\lambda_2^2\left(1+\kappa^2\right)\right]\kappa CS.
\end{align}
We may then compute the corresponding Cauchy stress components. 
It follows from \eqref{ssa} that  $T_{13}=T_{23}=0$. 
We complete the plane stress assumption by setting $T_{33}=0$, which gives us the expression for $p$ and the  in-plane stress components are therefore \cite{Cism, HaO}
\begin{align} \label{ips}
& T_{11}  =2W_1\left(B_{11}-B_{33}\right)+2W_2\left[B_{22}\left(B_{11}-B_{33}\right)-B_{12}^2\right]+2W_4a_1^2+4W_5a_1(Ba)_1, \nonumber \\
& T_{22}  =2W_1\left(B_{22}-B_{33}\right)+2W_2\left[B_{11}\left(B_{22}-B_{33}\right)-B_{12}^2\right]+ 2W_4a_2^2+4W_5a_2(Ba)_2, \nonumber \\
& T_{12}  = 2W_1B_{12}+2W_2B_{12}B_{33}+2W_4a_1a_2+2W_5\left[a_1(Ba)_2+a_2(Ba)_1\right],
\end{align}
where we used $\det \vec B = 1$ (incompressibility) to compute the components of $\vec B^{-1}$.
Here,  $a_i$ and $(Ba)_i$ denote the appropriate components of the vectors $\vec a$ and $\vec{Ba}$, respectively.
Explicitly, they read
\begin{align} \label{basim}
& a_1= \lambda_1C + \lambda_2 \kappa S, && a_2= \lambda_2 S, \nonumber \\
&(Ba)_1= \lambda_1\left(\lambda_1^2 + \lambda_2^2 \kappa^2 \right)C + \lambda_2\left(\lambda_1^2+\lambda_2^2+\kappa^2\lambda_2^2\right)\kappa S, 
 && (Ba)_2 = \lambda_1\lambda_2^2 \kappa C + \lambda_2^3\left(1+\kappa^2\right)S.
\end{align}
We remark that $a_2=\lambda_2 \sin \Phi \ne 0$, and thus deformed fibres are \emph{never}  aligned with the direction of the applied force.

For tensile testing,
\be \label{steo}
T_{22}=T \ne 0, \qquad T_{11}=T_{12}=0.
\en
Since two of the in-plane stresses are identically zero, it follows from \eqref{ips} that for the classes of materials that depend on only \emph{two} invariants a relationship between the deformation parameters $\lambda_1,\lambda_2,\kappa$ will be obatined.
For a given general fibre-angle $\Phi \, (\ne 0, \pi/2$), this relation therefore reduces the number of independent kinematical variables by one.  In contrast to the relations that result from imposing the physical constraints of, say, incompressibility and inextensibility, there is no physical motivation for these restrictions;  these relations are merely the result of a constitutive choice. 

As an example, consider strain-energy functions of the form \eqref{sepw}, which are almost universally used when modelling transversely isotropic materials, including biological, soft tissue. 
This choice is motivated purely by mathematical convenience, with two invariants being the minimum necessary to include both an isotropic and an anisotropic contribution to the strain-energy function and the $\left(I_1, I_4\right)$ pair being chosen because the resulting form of the stress-strain relation \eqref{ssa} is particularly convenient.  
For strain-energy functions of the form \eqref{sepw}, the simultaneous satisfaction of  \eqref{steo}$_{2,3}$ yields the following linear, homogeneous system of two equations for $W_1$ and $W_4$:
$$
0=W_1\left(B_{11}-B_{33}\right)+W_4a_1^2, \quad 0 = W_1B_{12}+W_4a_1a_2,
$$
 which, since $a_1 \ne 0$, gives non-trivial solutions for $W_1,W_4$ if, and only if, the following, \emph{purely kinematical}, relation holds:
\be \label{ui1i4}
 \lambda_1 \left(1 - \lambda_1^{-4}\lambda_2^{-2}\right)S = \lambda_2 \kappa C.
\en
This relationship is valid \emph{for all materials for which $W=W(I_1,I_4)$}.
It is therefore a necessary test of this constitutive hypothesis; if for \emph{any} non-zero angle of orientation \eqref{ui1i4} is violated \emph{at any stage} during simple tension,  then $W$ is \emph{not} a function of $I_1$ and $I_4$ only. It is shown in the next section that this kinematical relation does not fit the data obtained from the simple tension of a composite consisting of a soft non-linear matrix reinforced with stiff linear fibres. 

Other pairs of invariants could be considered in the same way as $\left(I_1,I_4\right)$. 
For example, if the strain energy density is chosen to depend on  $I_2$ and $I_4$ only, then the corresponding semi-universal relation has the form
\be
 \left(\lambda_1^4\lambda_2^2 - 1\right)\lambda_2S = \kappa\left(\lambda_1C + \lambda_2 \kappa S \right).
\en
Similar considerations apply for the pairs $(I_1,I_5)$ and $(I_2,I_5)$. 

\begin{figure} [htp!]
\begin{center} \epsfig{width=0.4\textwidth, figure=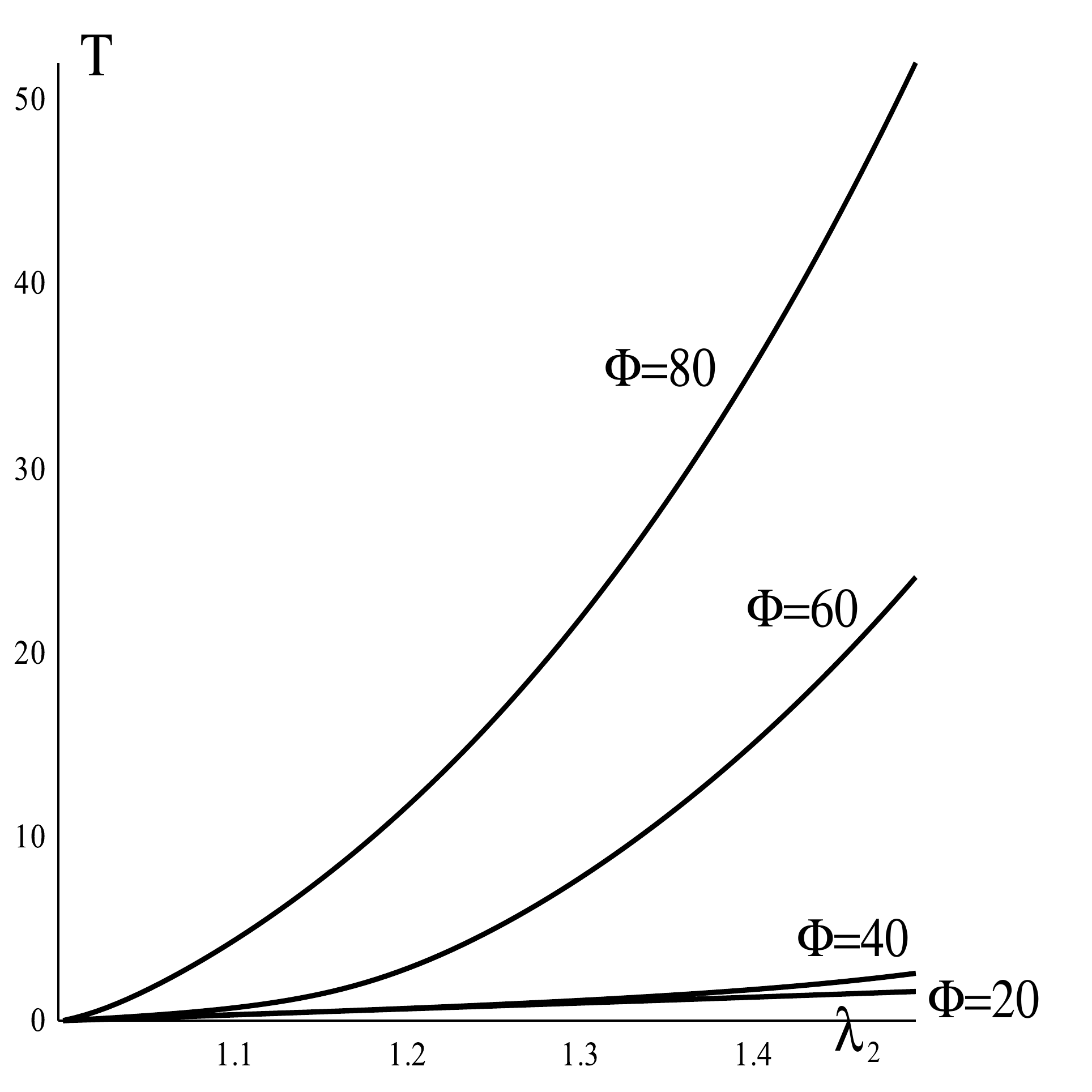}\end{center}
\caption{{\small Tensile stress vs tensile stretch for the standard reinforcing material. Here the stress measure is normalised with respect to the base shear modulus, $T \equiv T_{22}/\mu$. The stretch $\lambda_2$ is measured in the direction of the tensile force, but is not the largest stretch in the sample because of the fibre reinforcement, see Fig.\ref{figEXP}. Here the fibers were at a angle $\Phi = 20^\circ, 40^\circ, 60^\circ, 80^\circ$ with respect to the normal to the tensile force in the reference configuration. }}
\label{figSRM}
\end{figure}

Assume for the moment that \eqref{ui1i4} holds for all members of the popular family of strain energies \eqref{sepw}. Then since $W_2=W_5=0$, it follows from \eqref{matrices}, \eqref{ips} that simple tension for these materials is described by the following two simultaneous equations in the two unknowns $\lambda_1, \lambda_2$:
\begin{align} \label{seq}
& T_{22} = 2\left(\lambda_2^2 - \lambda_1^{-2}\lambda_2^{-2}\right)W_1 + 2 \lambda_2^2 S^2 W_4, \nonumber \\ 
&0 = \left(1 - \lambda_1^{-4}\lambda_2^{-2}\right) W_1 + \left(1 - \lambda_1^{-4}\lambda_2^{-2} S^2\right) W_4,
\end{align}
with $I_1$, $I_4$ now given by 
\be
I_1 = \dfrac{\lambda_1^2}{C^2} + \lambda_2^2 + \lambda_1^{-2}\lambda_2^{-2}\left(1-\dfrac{S^2}{C^2}\right),
\qquad
 I_4 = \dfrac{\lambda_1^2}{C^2}\left(1-\lambda_1^{-4} \lambda_2^{-2} S^2\right)^2 + \lambda_2^{2}S^2,
\en
where we eliminated $\kappa$ using \eqref{ui1i4}. For the particular example of  the standard reinforcing model \eqref{srm}, these two equations are
\begin{align} 
& T\equiv T_{22}/\mu = \lambda_2^2  - \lambda_1^{-2}\lambda_2^{-2}  + 2 \gamma\lambda_2^2 S^2 \left(I_4-1\right), 
\label{SRM1}\\[4pt]
& 0 = 1 - \lambda_1^{-4}\lambda_2^{-2} + 2\gamma \left(1 - \lambda_1^{-4}\lambda_2^{-2}S^2 \right) \left(I_4-1\right),
\label{SRM2}
\end{align}
with $I_4$ given just above.
These equations suggest a protocol to determine whether a given anisotropic soft tissue can be modelled by the standard reinforcing material, once it has been established that its strain-energy density is of the form \eqref{sepw} by first checking experimentally  that the semi-universal relation \eqref{ui1i4} is satisfied.
First, plot $1 - \lambda_1^{-4}\lambda_2^{-2}$ against $-2\left(1 - \lambda_1^{-4}\lambda_2^{-2}S^2 \right) \left(I_4-1\right)$: if a linear regression analysis reveals that the relationship between the two quantities is linear (up to a certain degree of approximation), then the slope of the curve gives the value of $\gamma$.
Next, plot the $T_{22}$ data against $\lambda_2^2  - \lambda_1^{-2}\lambda_2^{-2}  + 2 \gamma\lambda_2^2 S^2 \left(I_4-1\right)$; if again, a linear relationship is found, then the material is adequately described by the standard reinforcing model, and the slope of that curve gives the value of $\mu$. Thus the experimental confirmation of the validity of the standard reinforcing model requires the satisfaction of at three demanding constitutive tests, given by \eqref{ui1i4},  \eqref{SRM1} and \eqref{SRM2}. 
To illustrate a typical tensile-stress-tensile stretch response for this model, we now fix $\gamma$ at $\gamma=10.0$, say, and vary the angle of the fibres $\Phi$ to produce Figure \ref{figSRM} in two steps: first, solve \eqref{SRM2} for a given $\lambda_2$ to find the corresponding $\lambda_1$; second, substitute into \eqref{SRM1} to find $T$.

\vspace{4 cm}
\noindent We see that, as expected intuitively, the more the fibers were oriented to be aligned with the direction of the tensile force (i.e. as $\Phi$ increases towards $90^\circ$), the stiffer the material response becomes.  Changing the value of $\gamma$ only brings quantitative changes but the trend remains the same. 
For a more complex strain-energy function, a multi-objective  optimization exercise must  take place in order to evaluate the material parameters \cite{OgSI04, niannaidh}.


\section{Simple tension test: numerical solution} \label{num}


A finite element model of a transversely isotropic block was built using ANSYS Version 13,  which allows reinforcing fibres to be randomly distributed throughout the matrix, as long as they are aligned in the same direction. If these fibres have identical material properties and orientation, then we may use a smeared reinforcement strategy to model the contribution of the fibres to the mechanical response of the fibre/matrix composite. The material parameters used in our simulations have a biomechanical motivation. Moulton \emph{et al.} \cite{Mea} found that the Young's modulus of passive myocardium is of the order of 0.02 MPa. The matrix was therefore assumed to be a neo-Hookean, non-linearly elastic material, since it is generally accepted that the neo-Hookean material is an excellent model of the mechanical response of general, \emph{isotropic} materials for strains of the order considered here (Yeoh and Fleming \cite{YaF}), with this value of Young's modulus.  The fibres are modelled as a relatively stiff linearly elastic material with a Young's modulus of 200 MPa, since Yamamoto \emph{et al.} \cite{Yea} found that the Young's modulus of collagen fascicles is of this order. In order to make the shearing component of the deformation clearly visible in  our graphics (see Figure \ref{45d} below), a volume fraction for the fibres of 0.1 was used; this fraction, however, is an order of magnitude greater than the volume fraction of interstitial collagen found in the heart (Van Kerckhoven \emph{et al.} \cite{coll}). The block had originally a width of 20mm, a height of 20mm and a thickness of 2mm. The nodes on the bottom surface of the block were constrained only so that no vertical displacement was allowed, simulating a clamp which is free to move laterally to allow for shearing of the specimen. No lateral displacement was allowed on the top surface of the block (again to simulate clamping), with a force acting in the positive vertical direction. The other surfaces of the block were assumed stress-free.

\begin{figure} [htp!]
\begin{center}
(a) 
\subfigure{\epsfig{width=0.40\textwidth, figure=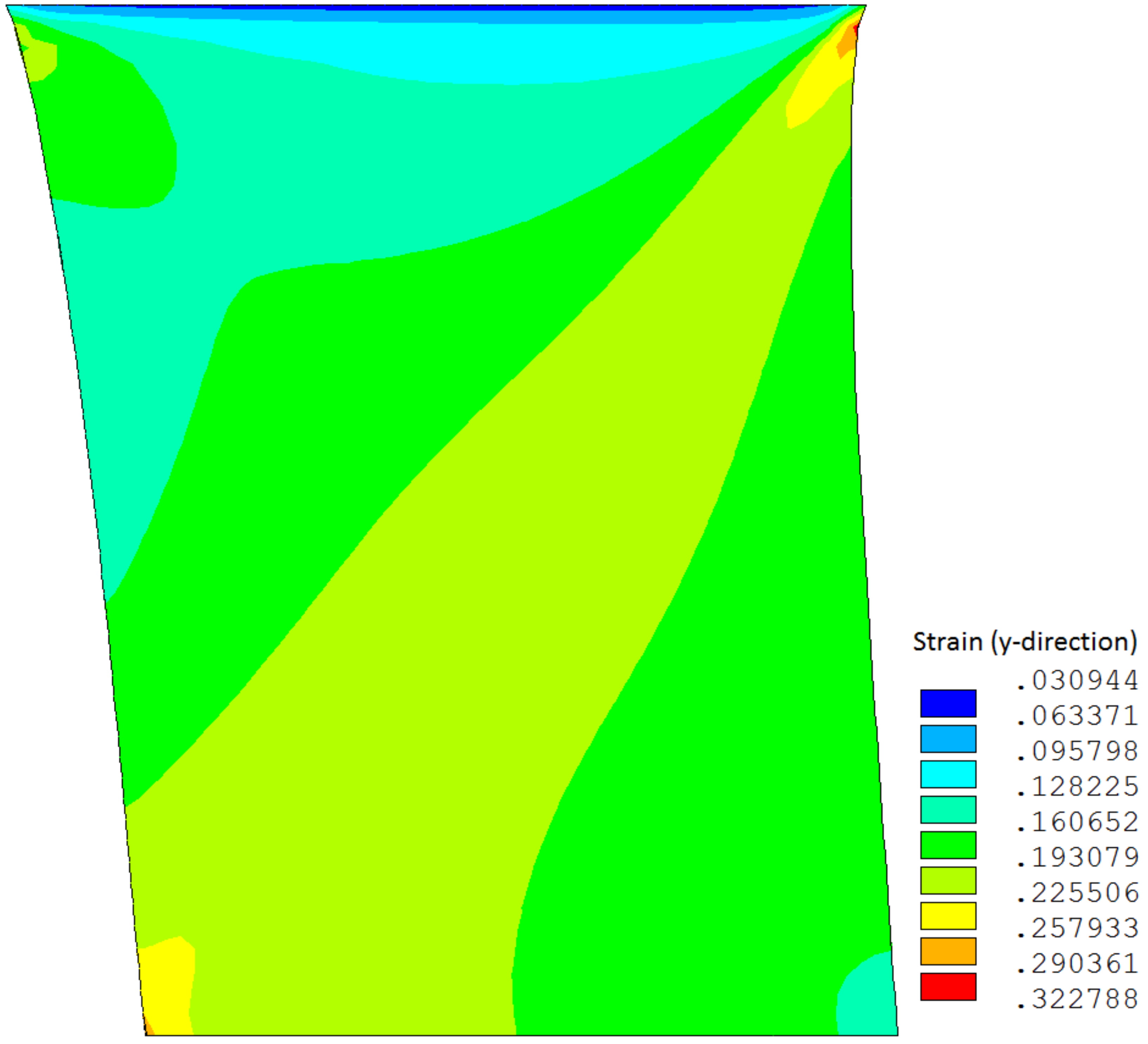}}
(b)
\subfigure{\epsfig{width=0.40\textwidth, figure=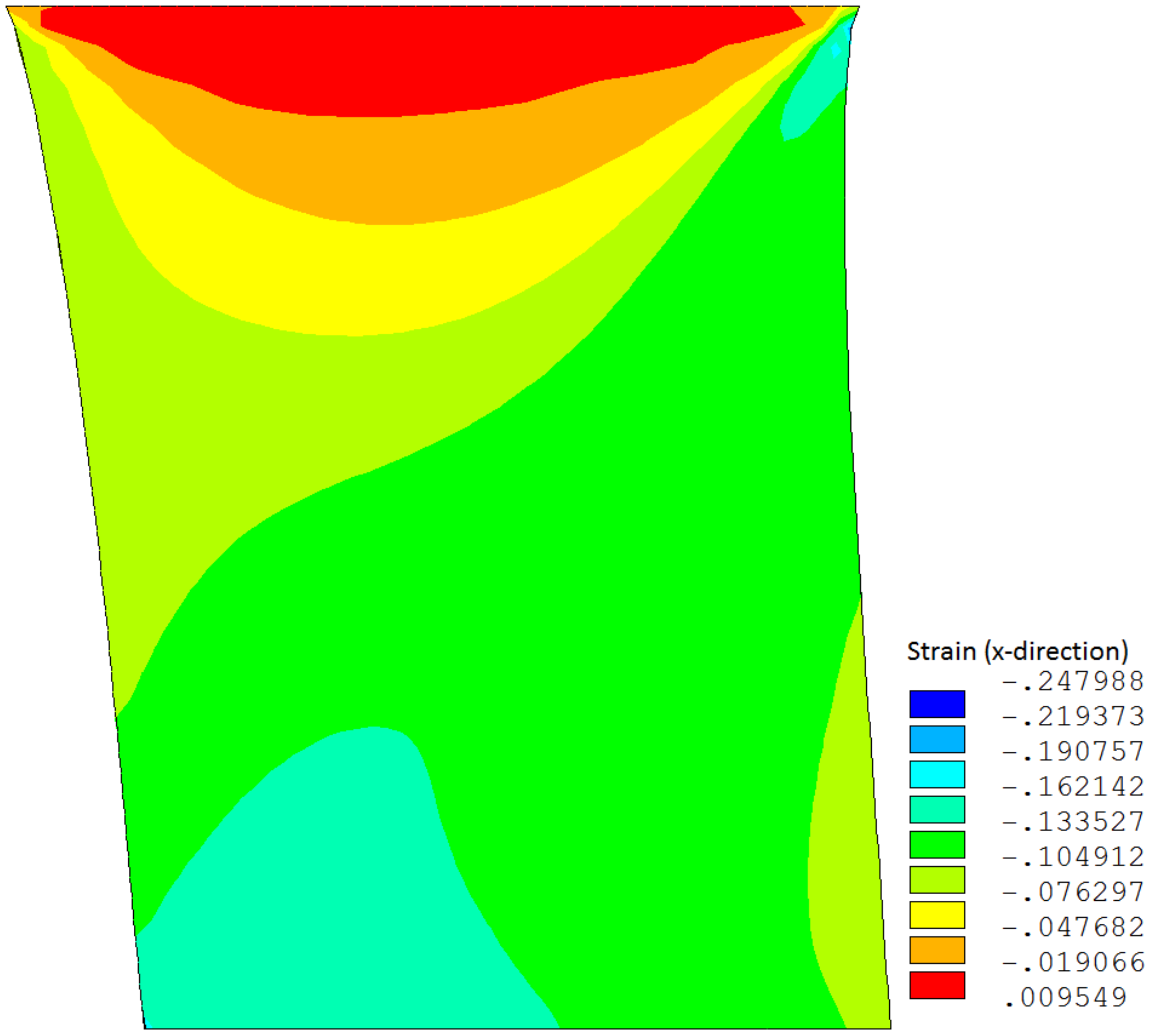}}\\
(c)
\subfigure{\epsfig{width=0.40\textwidth, figure=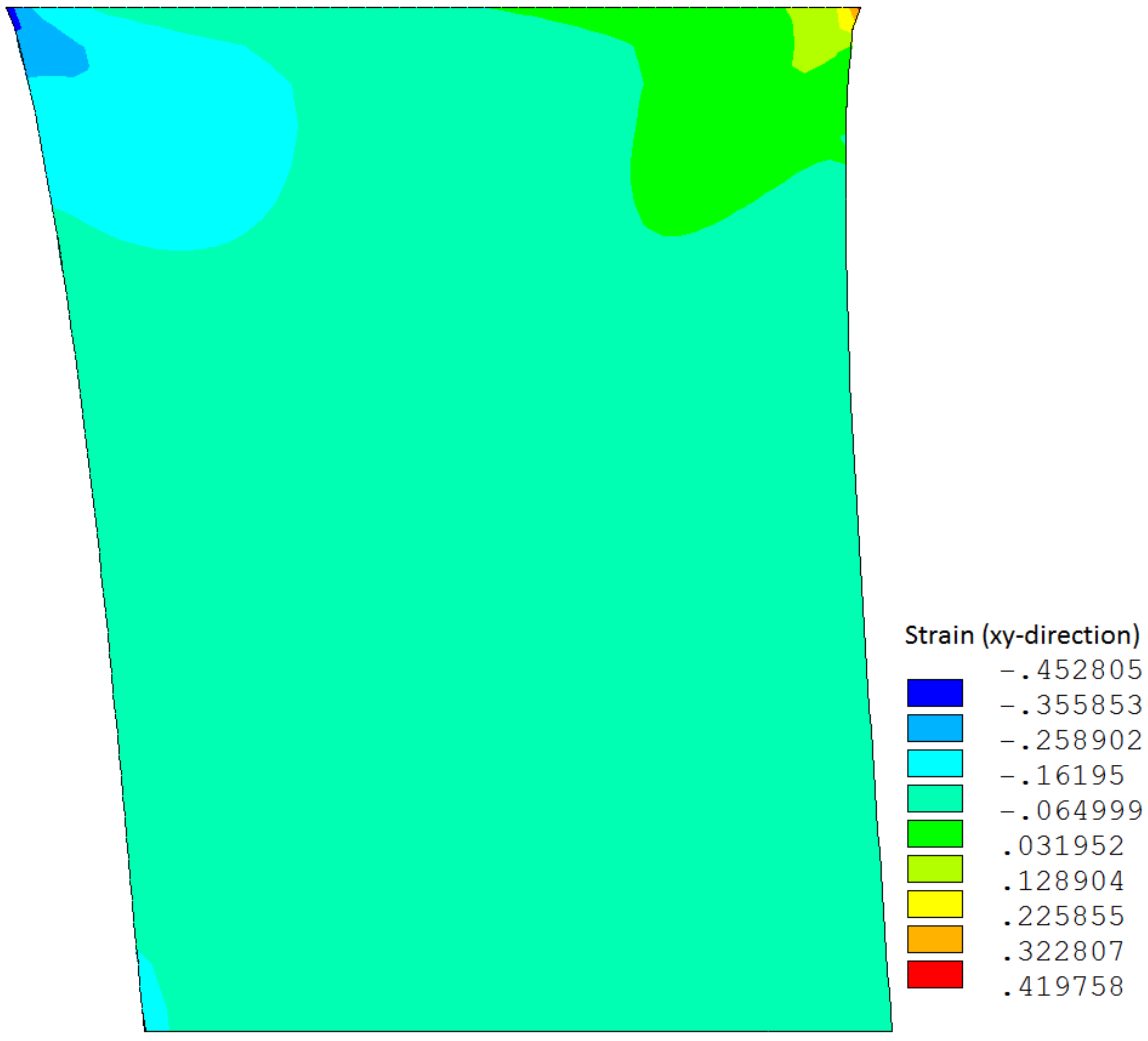}}
\end{center}
\caption{{\small Contour plots of (a) normal strain; (b) transverse strain; (c) shear strain}}
\label{contour}
\end{figure}

As an illustrative example of the Finite Element simulations conducted, the mid-plane of the thickness in the initial and deformed configurations is shown in Figure \ref{45d} for the fibre composite with $\Phi = 45^{\circ}$, subjected to an axial strain of $1.2$. For comparison purposes, these configurations are also given in Figures \ref{20}, \ref{80} in the Appendix for both  $\Phi = 20^{\circ}$, and  $\Phi = 80^{\circ}$. We note that the out-of-plane deformations in all our simulations were essentially homogenous, with inhomogeneity confined to thin boundary layer-like regions near the clamped ends. The corresponding contour plots of the axial, transverse and shear strains for $\Phi = 45^{\circ}$ are given in Figure \ref{contour}, so that the degree of homogeneity can be assessed. These graphics confirm our physical intuition that deformations of the form \eqref{def} are good models of the deformation that results from subjecting transversely isotropic materials to simple tension, especially through the central region of the specimen. It also supports our contention that an important application of the analysis presented here is that of modelling the mechanical response of biological, soft tissue given the excellent qualitative agreement between the edge profiles of the numerical simulations and the experimental results of, for example, Guo \emph{et al.} \cite{Guoetal}, who performed finite simple shear tests on porcine skin in order to obtain guidelines for the selection of specimen aspect ratio and clamping prestrain when studying the material response of soft tissues under simple-shear tests. Although the assumed homogeneous deformation \eqref{def} isn't an exact fit with the numerical results (the differing amounts through which the bottom corners of the specimen are sheared are testament to that), nonetheless the homogeneous approximation should be more than adequate for our constitutive modelling purposes.

\begin{figure} [htp!]
\begin{center} 
\subfigure{\epsfig{width=0.4\textwidth, figure=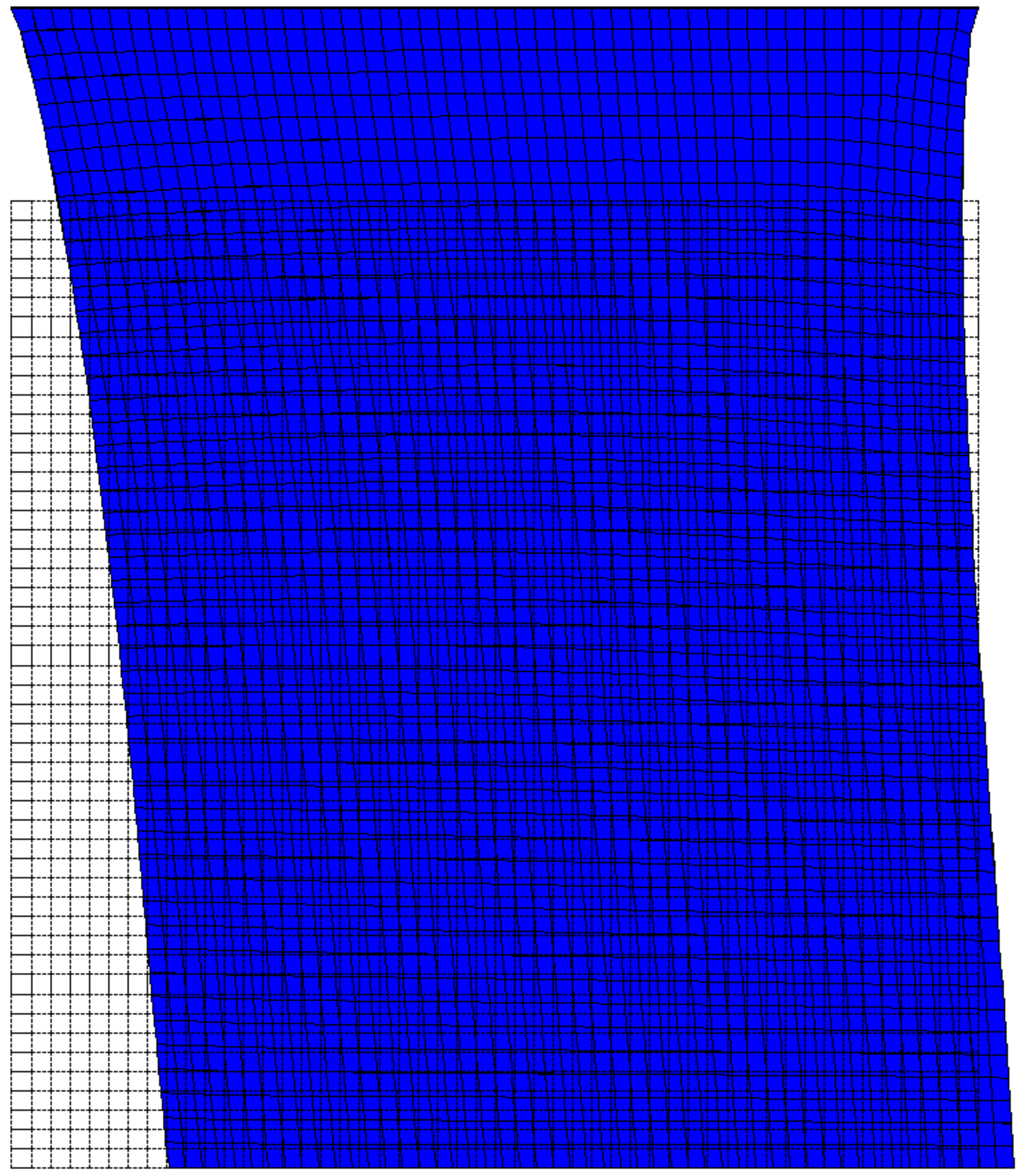}}
\end{center}
\caption{{\small Initial configuration and final configuration with $\lambda_2=1.2$ for $45^{\circ}$ initial fibre orientation.}}
\label{45d}
\end{figure}

Although the focus here is on simple tension testing, and therefore on stress controlled tests, for the numerical experiments it was convenient to control the axial stretch $\lambda_2$ so that strains consistent with the physiological regime (of the order of 20$\%$) were reproduced. Consequently axial stretches up to $\lambda_2 =1.2$ were imposed for a number of  different initial fibre orientations. All of our numerical results were qualitatively the same and the results for $\Phi = 45^{\circ}$ are taken as representative. The transverse stretches $\lambda_1$ and amounts of shear $\kappa$ for $\Phi = 45^{\circ}$ were calculated by measuring the displacement of the edge nodes along the centre of the specimen, where, as can be seen from Figures \ref{45d}, \ref{contour}, the end effects are minimised and the deformation is essentially homogeneous. 
The numerical results are given in the Appendix and are summarised in graphical form in Figure \ref{numres} below:

\begin{figure} [htp!]
\begin{center} 
\subfigure{\epsfig{width=0.75\textwidth, figure=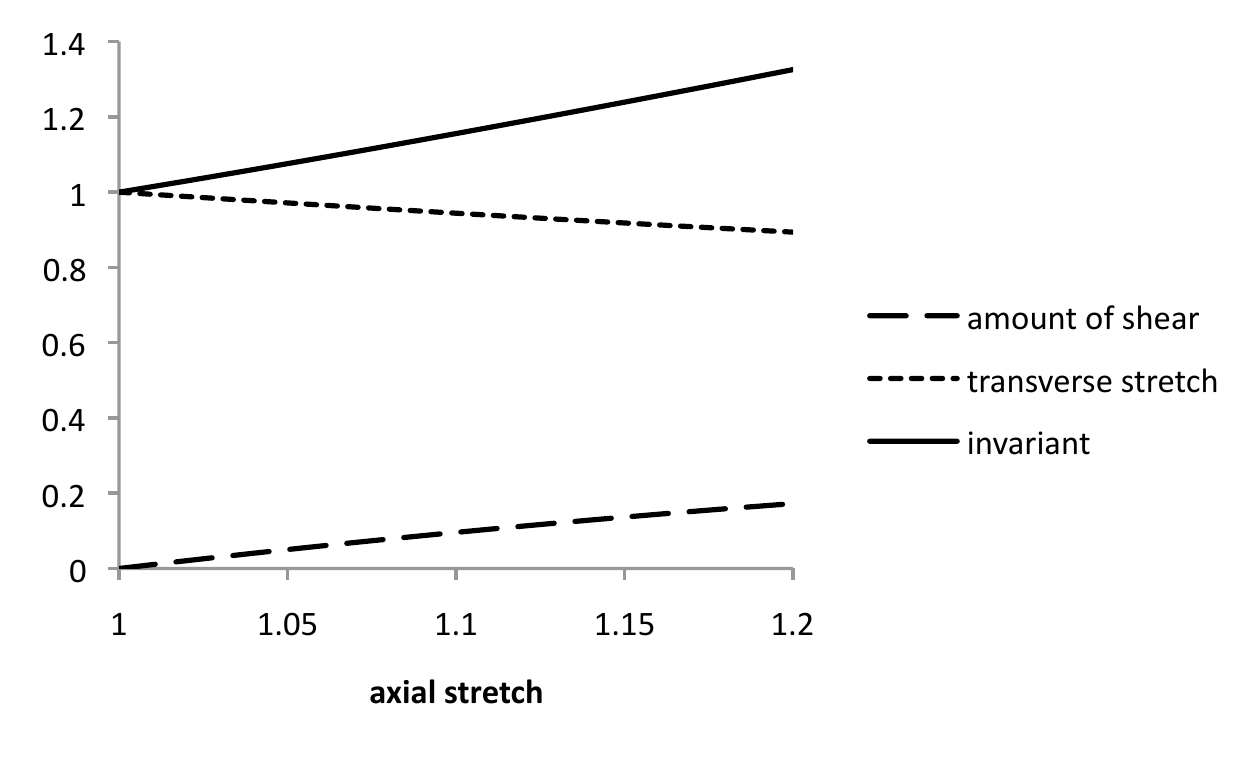}}
\end{center}
\caption{{\small Typical transverse stretch, shear and $I_4$ results.}}
\label{numres}
\end{figure}

\vspace{4 cm}
As might be expected, the transverse stretch is a monotonically decreasing and the amount of shear a monotonically increasing function of the imposed axial stretch.  
As a check of the above predictions, the invariant $I_4$ was computed using \eqref{i4}$_1$. Recalling that this invariant is the square of the fibre stretch, one would expect $I_4$ to increase with increasing axial stretch. This is reflected in Figure \ref{numres}, where the fibre stretch is always greater than one, thus avoiding possible instabilities arising from fibres being in compression.

\begin{figure} [htp!]
\begin{center}
(a) 
\subfigure{\epsfig{width=0.45\textwidth, figure=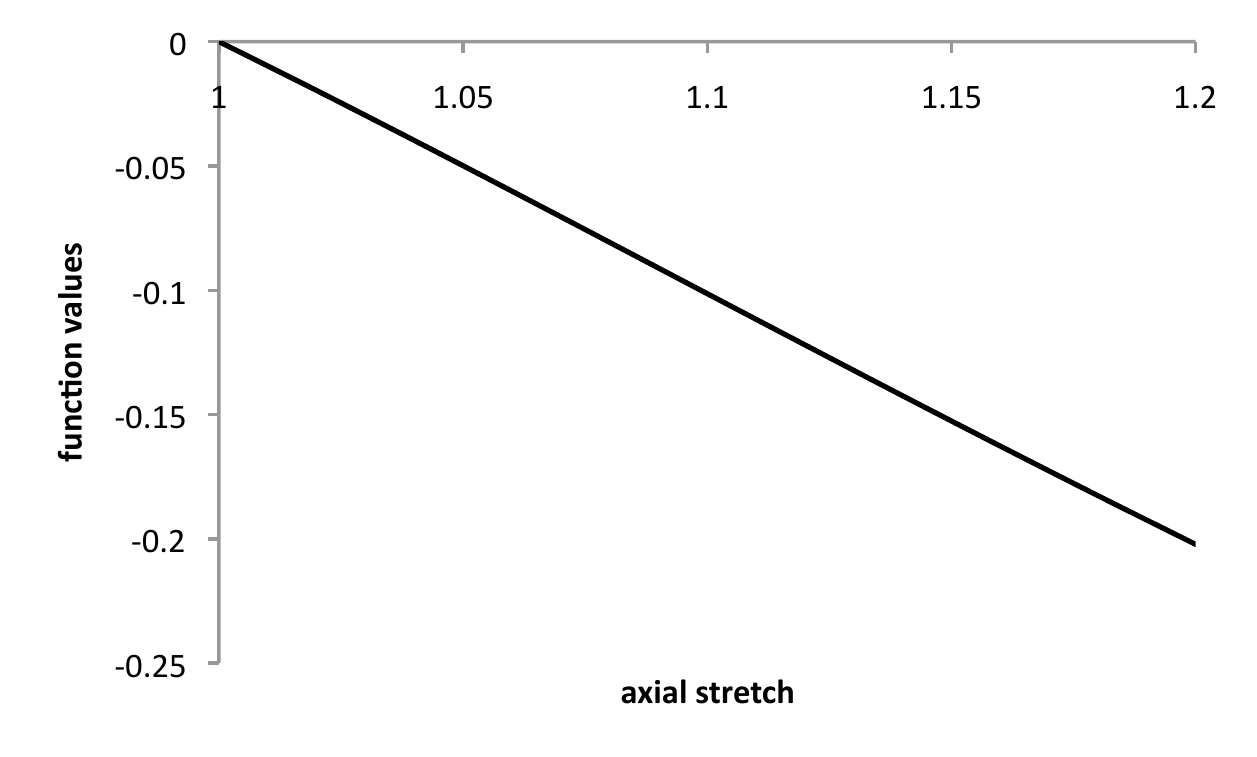}}
(b)
\subfigure{\epsfig{width=0.45\textwidth, figure=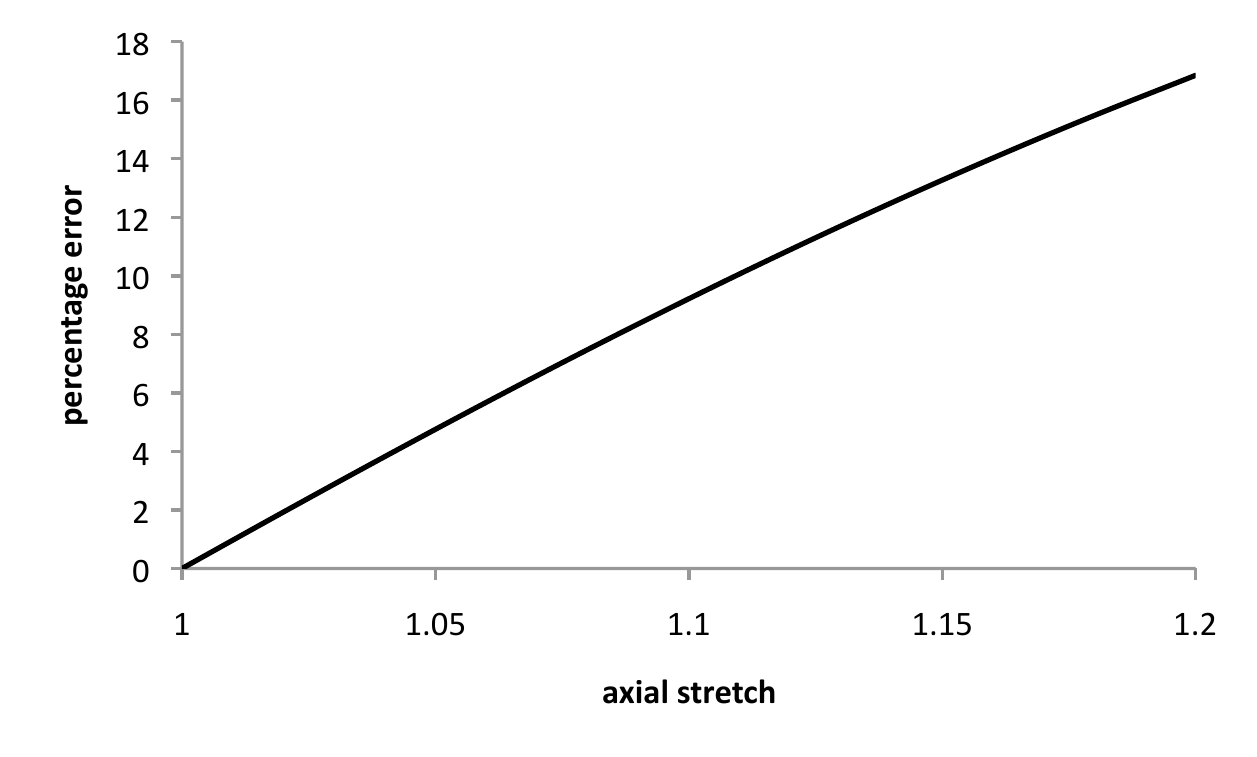}}
\end{center}
\caption{{\small  (a) Test of $\left(I_1,I_4\right)$ hypothesis; (b) Percentage error plots.}}
\label{fplot}
\end{figure}

The validity of kinematical relations like \eqref{ui1i4}, obtained by assuming invariant pairs for the strain-energy function, is now examined by comparing them with the numerical predictions of the behaviour of fibre-reinforced composites provided by our Finite Element simulations. 
Only the relation \eqref{ui1i4} is under consideration here, but similar results were obtained for the other possible kinematical relations. 
Strain controlled experiments were performed in the Finite Element analysis and we therefore consider $\lambda_1$ and $\kappa$ as functions of the axial stretch $\lambda_2$ and therefore let
\be \label{deff}
f\left(\lambda_2\right) \equiv  \lambda_1\left(1 - \lambda_1^{-4}\lambda_2^{-2}\right)S - \lambda_2 \kappa C,
\en 
where $\lambda_1 = \lambda_1(\lambda_2)$, $\kappa = \kappa(\lambda_2)$.
The analysis of Section \ref{st} has shown that if $W=W\left(I_1,I_4\right)$, then $f\left(\lambda_2\right)$ should be zero over the range of axial stretch of interest \emph{for all fibre angles}. 
For the physiological range of strain $1.0 \le \lambda_2 \le 1.2$, the function values for our illustrative fibre angle of  $\Phi=45^{\circ}$ are plotted in Figure \ref{fplot}(a). The practical difficulty in interpreting the data of Figure \ref{fplot}(a) is that a natural measure of `closeness' between the function \eqref{deff} and $0$ is not available. 
One possible solution is to normalise the absolute differences between the $f\left(\lambda_2\right)$ values and $0$ using the applied stretch $\lambda_2$ and then interpret the results as percentage errors. Plots of these errors, defined therefore by
\begin{equation}
\text{percentage error} \equiv  \left| \frac{f\left(\lambda_2\right)}{\lambda_2} \right| \times 100,
\end{equation}
are given in Figure \ref{fplot}(b).

The percentage error plot strongly suggests that the relation \eqref{ui1i4} is not valid for $\Phi = 45^{\circ}$ and thus we conclude that the numerical simulations are not supportive of the constitutive assumption $W=W\left(I_1,I_4\right)$, since if this assumption were true, then  \eqref{ui1i4} would hold for all fibre angles. Consequently, to model the full range of fibre orientations for physiological strains, the Finite Element simulations suggest that \emph{at least three invariants are required to fully capture the mechanical response of transversely isotropic materials}. It should be noted, however, that our simulations show a pronounced fibre-effect on the percentage error values: specifically, the percentage errors decrease with decreasing orientation angle. This is to be expected because in the limiting case of $\Phi=0^{\circ}$ the fibres are perpendicular to the direction of the applied force and consequently have no effect on the composite response to the applied stress distribution.

Holzapfel and Ogden \cite{HaO} considered the extent to which biaxial testing can be used to determine the elastic properties of transversely isotropic materials (the same problem for strain energies based on limited structural information and multiaxial stress-strain data was considered by Humphrey and Yin \cite{Hump1} and Humphrey \emph{et al.} \cite{Hump2}).  In particular, they concluded that \emph{if the constitutive assumption \eqref{sepw} is valid}, then biaxial tests can be used to determine the functions $W_1,W_4$ and hence to determine the form of the corresponding strain-energy function \eqref{sepw}.   Our conclusions do not contradict their results; rather our results cast doubt on their premise. It is our contention that strain-energy functions of the form \eqref{sepw} are \emph{not} valid.


\section{Tractions along the inclined faces} \label{inclined}


Although physical intuition and the numerical experiments of the Section \ref{num} suggest that the deformation \eqref{def} is likely to be an excellent approximation to the deformed state of a rectangular block subjected to the tension field \eqref{steo}, the semi-inverse approach adopted here results in an over-determined system for the unknowns $\lambda_1, \, \lambda_2, \, \kappa$,  with some boundary conditions in any physical realisation of the proposed experiments not being satisfied. Specifically, it is envisaged that the inclined faces of the specimen will be stress-free but, as is well-known for isotropic materials (see, for example, Atkin and Fox \cite{AaF} for a clear discussion of the issues involved), normal and shear stresses must be applied to the inclined faces of the block in order to maintain a state of homogeneous deformation.

\begin{figure} [htp!]
\begin{center} \epsfig{width=0.6\textwidth, figure=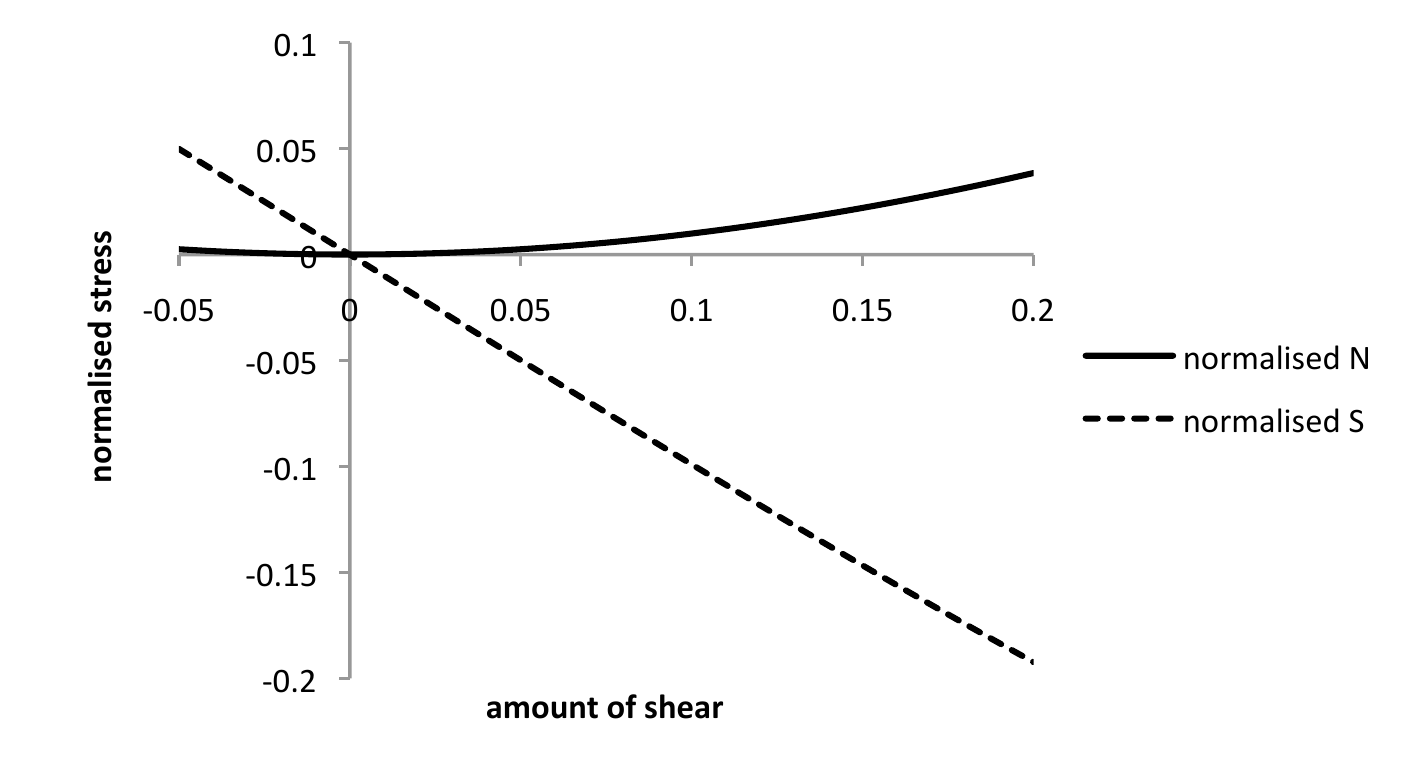}\end{center}
\caption{{\small  Stresses on inclined faces.}}
\label{NS}
\end{figure}

It is easily shown here that for the tensile test deformation, the outward unit normal $\vec n$ to the inclined faces in the deformed configuration has the following components:
\be
\vec{n}=\left(\frac{1}{\sqrt{1+\kappa^2}},-\frac{\kappa}{\sqrt{1+\kappa^2}},0\right),
\en
independent of the axial and lateral stretches.
Noting the imposed state of stress, \eqref{steo}, the normal stress $N$ and the shear stress $S$ that therefore have to applied to the inclined faces in order to maintain a block in the deformed state \eqref{def} are given by
\begin{equation}
N=\frac{\kappa^2}{1+\kappa^2}T, \qquad 
S=-\frac{\kappa}{1+\kappa^2}T.
\end{equation}
The normalised stresses $\hat{N} \equiv N/T$, $\hat{S} \equiv S/T$ are therefore a function only of the amount of shear $\kappa$ and are plotted in Figure \ref{NS} for the moderate range of $\kappa$ suggested by the simulations of the last section.

It is clear from this figure that compared to the tensile stress, only insignificant normal stresses are required for physiological strains; consequently the absence of normal stresses applied to the inclined faces is likely to have a negligible effect on the homogeneity of the deformation. 
In contrast, there is essentially a linear relationship between $\kappa$ and $\hat{S}$ over the the range of strain of interest. That relatively large shear stresses are required to maintain homogeneity is not surprising given that even the linear theory for isotropic materials requires a shear stress on the inclined face of equal magnitude to the shear stress driving the deformation. It is worth emphasising here, however, that for the given strain range, the shear stresses are essentially an order of magnitude smaller than the applied tensile stress.  
In practice a lack of shear stress on the inclined faces of sheared blocks of biological tissue does not seem to affect the homogeneity of the deformation. 
In Dokos \emph{et al.} \cite{Dokos}, for example, cuboid specimens of myocardium were sheared up to 40$\%$ with no reported mention of any inhomogeneity observed in testing. Some protocols to minimise inhomogeneity when shearing biological, soft tissue were proposed by Horgan and Murphy \cite{Hor}. Certainly for the experiments considered here, Figure \ref{contour} suggests that homogeneity is likely to be maintained, at least within the central region of the specimen.


\section{Conclusions}


A method has been proposed for the off-axis simple tension testing of transversely isotropic nonlinearly hyperelastic materials, a method that should be a viable alternative to the dominant biaxial tension test for material characterisation. This method proposes that a shearing deformation accompanies a triaxial stretching regime. 
It was shown that if, as is commonly the case, a pair of one isotropic and one anisotropic invariant is chosen as the basis for the strain-energy function, then a kinematical universal relation must be satisfied for this new testing regime, one that must hold for all fibre angles and for the full range of applied tension. Finite Element simulations suggest that this is too demanding a requirement and that at least three invariants are necessary to model the full range of mechanical response of transversely isotropic materials.


\section{Appendix}


\subsection{Numerical results}
The following procedure was adopted to calculate the kinematical quantities $\lambda_1,\lambda_2,\kappa$ from the numerical results. 
The $x_1$ coordinates of the mid-points of the inclined faces were isolated from the rest of the output data at specified values of the prescribed axial stretch; call them $x_1^l,x_1^r$, using an obvious notation. Referring to Figure \ref{45d}, let the origin coincide the bottom left corner of the undeformed block. Since the dimensions of the block were chosen to be 20mm $\times$ 20mm $\times$ 2mm, it follows from \eqref{def} that
\[
x_1^l=\kappa \lambda_2 10, \qquad x_1^r=\lambda_1 20+ \kappa \lambda_2 10.
\]
Since $\lambda_2$ is controlled, $\kappa, \lambda_1$ are therefore obtained from
\[
\kappa = \frac{x_1^l}{10\lambda_2}, \qquad \lambda_1=\frac{x_1^r-x_1^l}{20}.
\]
These calculated quantities to four decimal places are tabulated in Table 1.

\begin{table} [htp!]
\begin{center} 
\begin{tabular}{| c | c | c |} 
\hline 
axial stretch & amount of shear & transverse stretch \\ 
\hline 
1&0&1\\
1.02&0.0207&	0.9885\\
1.04&0.0406&	0.9771\\
1.06&0.0598&	0.9659\\
1.08&0.0782&	0.9549\\
1.1&0.0959&	0.9441\\
1.12&0.1128&	0.9336\\
1.14&0.1288&	0.9232\\
1.16&0.1442&	0.9132\\
1.18&0.1587&	0.9033\\
1.2&0.1725&	0.8938\\
\hline 
\end{tabular} 
\caption{Kinematics for numerical experiments with a $45^{\circ}$ fibre angle}
\end{center} 
\end{table}

\vspace{7 cm}

\subsection{Comparative plots}

\begin{figure} [htp!]
\begin{center} 
\subfigure{\epsfig{width=0.35\textwidth, figure=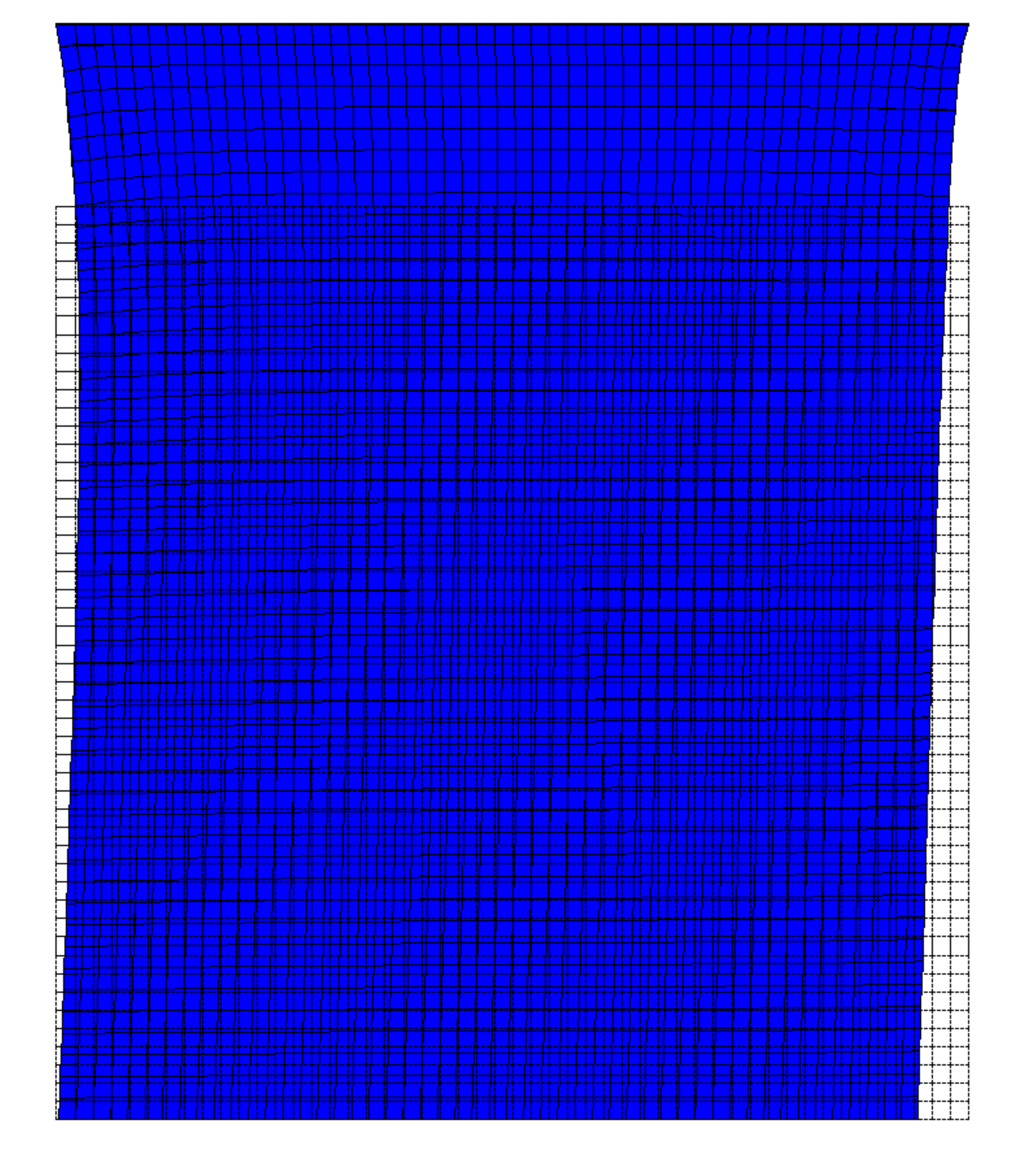}}
\end{center}
\caption{{\small Initial and final configurations with $\lambda_2=1.2$ for $20^{\circ}$ initial fibre orientation.}}
\label{20}
\end{figure}

\begin{figure} [htp!]
\begin{center} 
\subfigure{\epsfig{width=0.35\textwidth, figure=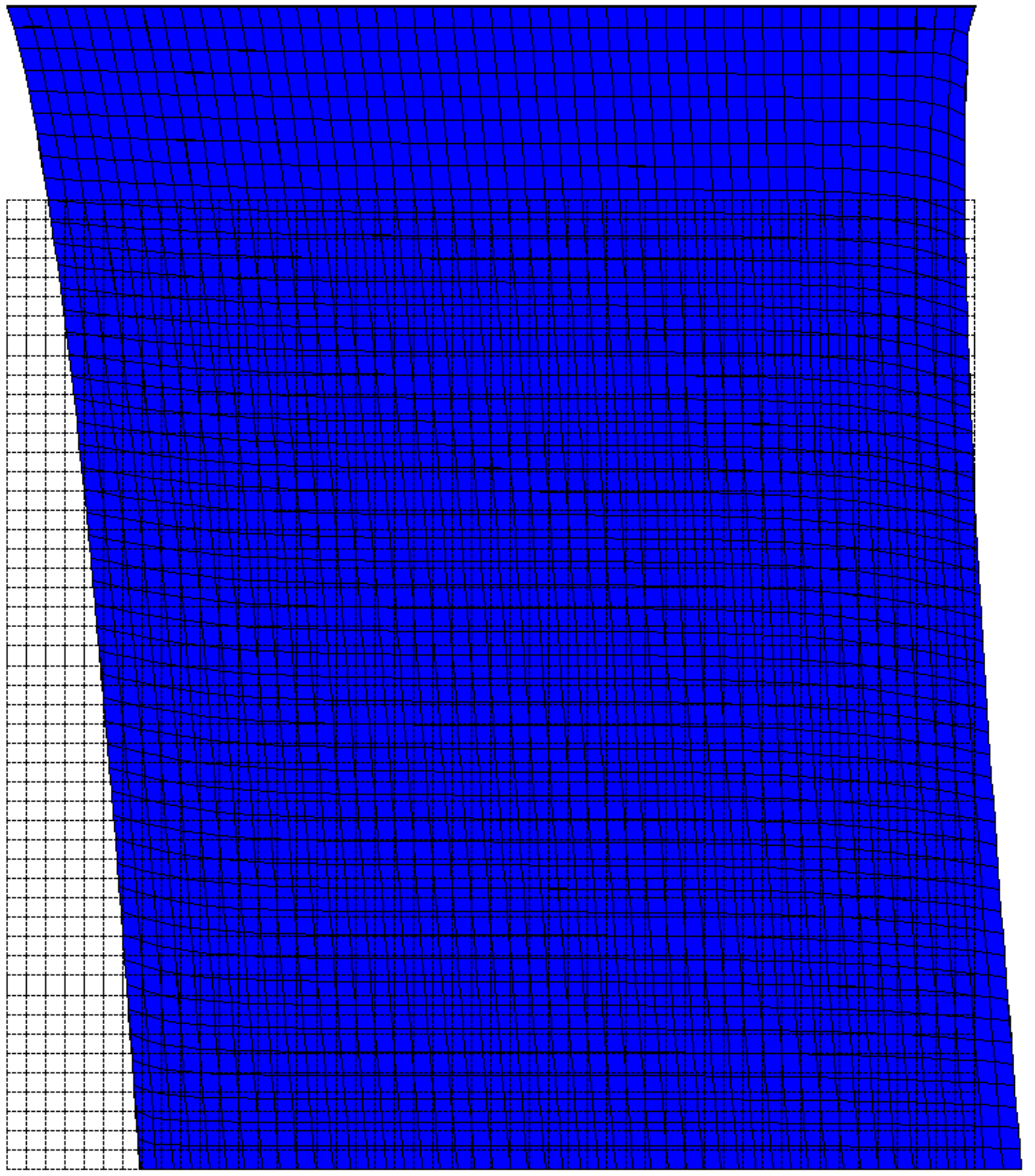}}
\end{center}
\caption{{\small Initial and final configurations with $\lambda_2=1.2$ for $80^{\circ}$ initial fibre orientation.}}
\label{80}
\end{figure}

Numerical results for a fibre angle of $45^{\circ}$ were presented in the main body of the paper. The simulations for this natural choice of fibre angle are supplemented below for an angle close to the horizontal and another close to the vertical.  For the material parameters used here, it is clear from Figures \ref{20}, \ref{80} that shear is negligible for a $20^{\circ}$ angle and much more pronounced for $80^{\circ}$. A comparison of these graphics with Figure \ref{45d} shows that the amount of shear for $45^{\circ}$ is between these two limiting cases. 

\vspace{1 cm}



\begin{thebibliography}{99}

\bibitem{Rivpap}
Rivlin, R.S. Collected Papers of R.S. Rivlin, vol. 1, Barenblatt, G.I., Joseph, D.D. (eds.). Springer, New
York (1997)

\bibitem{Spen}
Spencer, A.J.M., 1972. 
Deformations of Fibre-Reinforced Materials. 
Oxford University Press.

\bibitem{PaH}
N.J. Pagano and J.C. Halpin, 1968.
 Influence of end constraint in the testing of anisotropic bodies. J. Compos. Mat. 2, 18--31.
 
 \bibitem{Marin}
 J.C. Mar\'{i}n, J. Ca\~{n}as, F. Par\'{i}s and J. Morton, 2002.
 Determination of G$_{12}$ by means of the off-axis tension test. Part I: review of gripping systems and correction factors. Composites: Part A 33, 87--100.
 
 \bibitem{Xea}
 Y. Xiao, M. Kawai and H. Hatta, 2010.
 An integrated method for off-axis tension and compression testing of unidirectional composites. J. Compos. Mat. 45, 657--669.

\bibitem{Moon}
H. Moon, C. Truesdell, 1974.
Interpretation of adscititious inequalities through the effects pure shear stress produces upon an isotropic elastic solid.
Arch. Rat. Mech. Analysis 55, 1--17.

\bibitem{RaW}
K.R. Rajagopal and  A.S. Wineman, 1987.
New universal relations for nonlinear isotropic elastic materials, J. Elasticity 17, 75--83. 

\bibitem{Goriely}
L.A. Mihai, A. Goriely, 2012.
Positive or negative Poynting effect? The role of adscititious inequalities in hyperelastic materials.
Proc. Roy. Soc. A (to appear).

\bibitem{Dea}
M. Destrade, J.G. Murphy and G. Saccomandi, 2012.
Simple shear is not so simple. International Journal of Nonlinear Mechanics 47, 210--214.

\bibitem{Rivlin}
R.S. Rivlin, 1948. 
Large elastic deformations of isotropic materials. IV. Further developments of the general theory. Philos. Trans. R. Soc. Lond. A 241, 379--397.

\bibitem{Gent}
A.N. Gent, J.B. Suh and S.G. Kelly III, 2007
Mechanics of rubber shear springs. Int. J. Nonlinear Mech. 42, 241--249.

\bibitem{HaM}
C.O. Horgan and J.G. Murphy, 2010.
Simple shearing of incompressible and slightly compressible isotropic nonlinearly elastic materials. 
Journal of Elasticity 98, 205--221.

\bibitem{Hump1}
J.D. Humphrey, F.C.P. Yin, 1987. 
A new constitutive formulation for characterizing the mechanical behavior of soft tissues. 
Biophysical Journal 52, 563-570.

\bibitem{Hump2}
J.D. Humphrey, R.K. Strumpf, F.C.P. Yin, 1990. 
Determination of a Constitutive Relation for Passive Myocardium: I. A New Functional
Form
J. Biomechanical Eng. 112, 333-339.

\bibitem{tor}
C.O. Horgan, J.G. Murphy, 2012. On the modeling of extension-torsion experimental data for transversely isotropic biological soft tissues, Journal of Elasticity 108, 179-191.

\bibitem{other}
J. F. Wenk, M. B. Ratcliffe, J. M. Guccione, 2012. Finite element modeling of mitral leaflet tissue using a layered shell approximation. Med. Biol. Eng. Comput. 50, 1071-1079.

\bibitem{Ning}
X. Ning, Q. Zhu, Y. Lanir, S.S. Margulies 2006. A transversely isotropic viscoelastic constitutive equation for brainstem undergoing finite deformation. Journal of Biomechanical Engineering 128, 925-933. 

\bibitem{Dest}
M. Destrade, M.D. Gilchrist, D.A. Prikazchikov, G. Saccomandi 2008.  Surface instability of sheared soft tissues. Journal of Biomechanical Engineering 130,  0610071-0610076. 



\bibitem{holz1}
G.A. Holzapfel, T.C. Gasser, R.W. Ogden, 2000. 
A new constitutive framework for arterial wall mechanics and a comparative study of material models. 
J. Elast. 61, 1--48.

\bibitem{holz2}
G.A. Holzapfel, R.W. Ogden, T.C. Gasser,  2006. 
Hyperelastic modelling of arterial layers with distributed collagen fibre orientations.
J. R. Soc. Interface 3, 15--35.

\bibitem{Cism}
Ogden, R.W., 2003. Nonlinear elasticity, anisotropy, material stability and residual stresses in soft tissue, In Biomechanics of Soft Tissue in Cardiovascular Systems, CISM Courses and Lectures Series no. 441, 65--108, Springer, Wien.



\bibitem{HaO}
G.A. Holzapfel, R.W. Ogden, 2009.
On planar biaxial tests for anisotropic nonlinearly elastic solids. A continuum mechanical framework. 
Mathematics and Mechanics of Solids 14, 474--489.



\bibitem{OgSI04}
R. W. Ogden, G. Saccomandi, I. Sgura, 2004.
Fitting hyperelastic models to experimental data.
Computational Mechanics 34, 484--502.

\bibitem{niannaidh}
A. N\`i Annaidh, K. Bruy\`ere, M. Destrade, M.D. Gilchrist, C. Maurini, M. Ott\'enio, G. Saccomandi, 2012.
Automated estimation of collagen fibre dispersion in the dermis and its contribution to the anisotropic behaviour of skin,
Annals Biomed. Eng. 40, 1666--1678. 

\bibitem{Mea}
M.J. Moulton, L.L. Creswell, R.L. Actis, K.W. Myers, M.W. Vannier, B.A. Szabo, M.K. Pasque,  1995. 
An inverse approach to determining myocardial material properties. 
J. Biomech. 28, 935--948.



\bibitem{YaF}
Yeoh, O.H., Fleming, P.D., A new attempt to reconcile the statistical and phenomenological theories of rubber elasticity, Journal of Polymer Science: Part B: Polymer Physics, Vol. 35, 1919-1931, 1997.

\bibitem{Yea}
Yamamoto, E., Hayashi, K., Yamamoto, N., 1999. 
Mechanical properties of collagen fascicles from the rabbit patellar tendon.  
J. Biomech. Eng. 121, 124--131.

\bibitem{coll}
Van Kerckhoven ,R., Kalkman, E.A.J., Saxena, P.R., Schoemaker, R.G., 2000. Altered cardiac collagen and associated changes in diastolic function of infarcted rat hearts. Cardiovascular Research 46, 316-323.




\bibitem{Guoetal}
Guo, D.-L., Chen, B.-S., Liou, N.-S., 2007.
 Investigating full-field deformation of planar soft tissue under simple-shear tests, Journal of Biomechanics 40,  1165-1170. 

\bibitem{AaF}
R.J. Atkin and N. Fox, 1980.
An introduction to the theory of elasticity, Longman, London.

\bibitem{Dokos}
Dokos, S., Smaill, B. H., Young, A. A. and LeGrice, I. J. 2002 
Shear properties of passive ventricular myocardium. Am. J. Physiol. Heart Circ. Physiol. 283, H2650-H2659.

\bibitem{Hor}
C.O. Horgan and J.G. Murphy, 2011
Simple shearing of soft biological tissues,  Proceedings of the Royal Society A 467, 760-777.

\end{thebibliography}
\end{document}